\documentclass[11pt]{article}

\usepackage{url}
\usepackage{graphicx}
\usepackage{amsmath}
\usepackage{amssymb}

\usepackage{makeidx}

\makeindex

\allowdisplaybreaks[2]

\begin{document}

\section{Gauge links, TMD-factorization, and TMD-factorization breaking}
\label{sec:mulders-rogers}


\hspace{\parindent}\parbox{\textwidth}{\slshape 
  Piet J. Mulders$\,^1$, Ted C. Rogers$\,^1$ \\[1ex]
$^1$\ Department of Physics and Astronomy, VU University, Amsterdam, Netherlands
}

\index{Mulders, Piet J.}
\index{Rogers, Ted}

\vspace{2\baselineskip}


\centerline{\bf Abstract}
\vspace{0.7\baselineskip}
\parbox{0.9\textwidth}{In this section, we discuss some basic
features of transverse momentum dependent, or unintegrated, 
parton distribution
functions. In particular, when these 
correlation functions are combined in a factorization formulae
with hard processes beyond the simplest cases, there are basic
problems with universality and factorization. We discuss some of these
problems as well as the opportunities that they offer.
}


\subsection{Introduction}

In hard processes, parton distribution functions (PDFs) and fragmentation functions (FFs) are
expressed as matrix elements of nonlocal combinations of quark or
gluon fields. In the collinear situation that all transverse momenta
of partons are integrated over in the definitions, the nonlocality is 
in essence light-like.
These correlation 
functions are convoluted with the squared amplitude for the partonic subprocess
(in essence the partonic cross section) of a hard process.  When the transverse momenta
of partons are involved, the non-locality in the matrix elements
includes a transverse separation, and a \emph{transverse momentum dependent} (TMD) factorization theorem is needed.
In all cases the definitions of the non-perturbative functions include gluon
contributions resummed into gauge links that bridge the nonlocality.

It is important to realize that the appearance of the gauge links 
is a consequence of the systematic resummation of extra gluon contributions 
in the derivations of factorization, so their structure is 
dictated by the requirements of factorization.

In processes like $\ell + H \longrightarrow \ell^\prime + h + X$
(semi-inclusive deep inelastic scattering), $\ell + \bar\ell \longrightarrow h_1 + h_2 + X$ 
(annihilation process) or $H_1 + H_2 \longrightarrow \ell + \bar \ell + X$
(Drell-Yan process) one has, 
at leading power in the hard scale, a simple underlying hard
process, which is a virtual photon (or weak boson) coupling to a
parton line. The color flow from hard part to 
collinear or soft parts is simple.
Additional gluons with polarizations collinear to the parton momenta
are resummed into gauge links, which exhibit the interesting behavior
that for transverse momentum dependent functions they bridge the 
transverse separation between the non-local field combinations at 
lightcone past or future infinity. Which gauge link is relevant in
a particular non-perturbative function depends on the color flow 
in the full process. For a quark distribution
function one has a link via (future) lightcone $+\infty$ if the color
flows into the final state, and a link via (past) lightcone $-\infty$
if the color is annihilated by another incoming parton.

QCD factorization theorems are central to understanding 
high energy hadronic scattering cross sections
in terms of the fundamentals of perturbative QCD.
In addition to providing a practical prescription for order-by-order calculations,
derivations of factorization provide a solid theoretical underpinning for  
concepts like PDFs and FFs which 
are crucial in the quest to expand the basic understanding of hadronic structure.
The most natural first attempt at a TMD-factorization formula is simply 
to extend the classic parton model intuition familiar from collinear factorization.
For the semi-inclusive deep inelastic scattering (SIDIS) cross section, for example, the cross section 
might be written schematically as
\begin{equation}
\label{eq:basic}
d \sigma \sim | \mathcal{H} |^2 \otimes \Phi(x,{\bf k}_T) \otimes \Delta(z,{\bf k}^\prime_T) \; \delta^{(2)} ({\bf q}_T + {\bf k}_T - {\bf k}^\prime_T).
\end{equation}
Here $\Phi(x,{\bf k}_T)$ is the TMD PDF while 
$\Delta(z,{\bf k}^\prime_T)$ is the TMD FF, with the usual probability 
interpretations, and $| \mathcal{H} |^2$ represents the hard part. 
The momentum ${\bf q}_T$ is the small 
momentum sensitive to intrinsic transverse momenta, ${\bf k}_T$ and ${\bf k}^\prime_T$, carried by the colliding proton and the produced hadron. 
The $\otimes$ symbol denotes all relevant convolution integrals, 
and the $x$ and $z$ arguments are the usual longitudinal momentum fractions.

In a perturbative derivation of factorization, a small-coupling 
perturbative expansion of the cross section is analyzed
in terms of ``leading regions'', and the sum is shown order-by-order 
to separate into the factors of equation~\eqref{eq:basic}.
The precise field theoretic definitions of the correlation 
functions, $ \Phi(x,{\bf k}_T)$ and $\Delta(z,{\bf k}^\prime_T)$, 
should emerge naturally from the requirements of factorization.  
In the hard part $| \mathcal{H} |^2$, all propagators must be off-shell 
by order the hard scale $Q$ so that asymptotic freedom 
applies, and small-coupling
perturbation theory is valid, with non-factorizing 
higher-twist contributions suppressed by powers of $Q$.  
Such factorization theorems are well-established for inclusive 
processes that utilize the standard \emph{integrated} 
correlation functions (see~\cite{Collins:1989gx} and 
references therein), but TMD-factorization theorems involve other subtleties, 
particularly with regard to the definitions of the 
of the TMD PDFs and FFs and their associated gauge links (or Wilson lines).
We will discuss some of these issues in the next few sections.

In cases where there is a more complex color flow such as is
often
the case
when the underlying hard process involves multiple color flows and/or
if the incoming partons are gluons, this 
can potentially lead to a more complex
gauge link structure including traced closed loops or looping gauge links.
For situations in which only one 
TMD correlation function 
is studied, these structures have been examined in~\cite{gaugelinks} for 
two-to-two partonic subprocesses.

In situations that involve several TMD functions, 
factorization
using seperate TMD functions fails completely, 
as we will discuss in the last section.

\section{Review: Collinear Factorization and Simple Gauge Links}
\label{sec:defs}
\begin{figure*}
\centering
  \begin{tabular}{c@{\hspace*{5mm}}c}
    \includegraphics[scale=0.6]{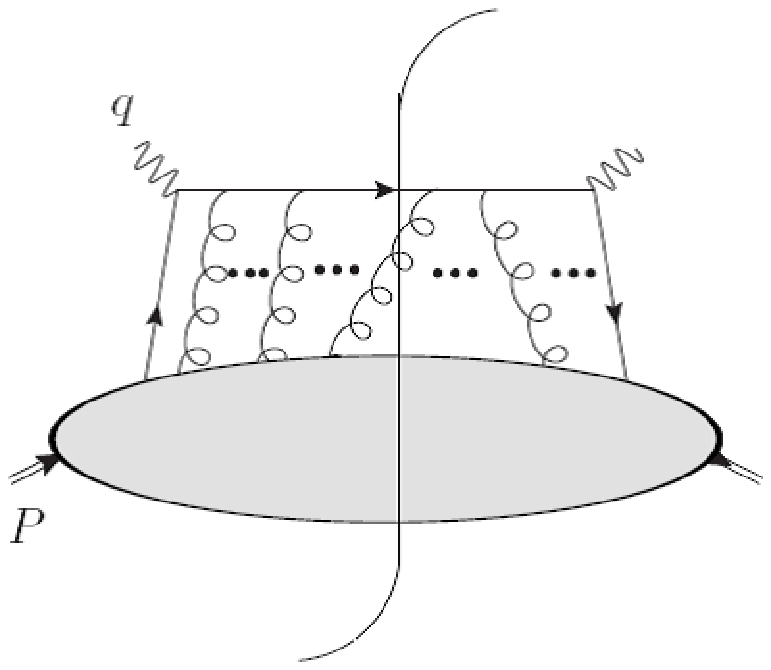}
    &
    \includegraphics[scale=0.6]{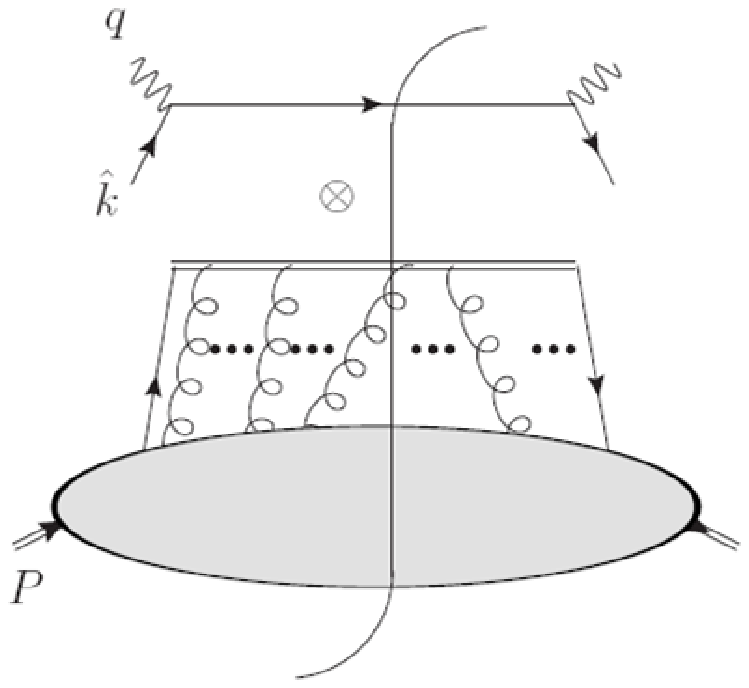}
  \\
  (a) & (b)
  \end{tabular}
\caption{(a) Target-collinear gluons in a graph for SIDIS.  (b) Factorization of extra gluons into 
gauge link contributions.}
\label{fig:basicfact}
\end{figure*}

To understand the issues that arise in defining TMDs, it is instructive to start 
with a review of the definition of the standard \emph{integrated} quark PDF
of collinear factorizaton.  It is
\begin{equation}
\label{eq:intDEF}
f(x;\mu) = {\rm F.T.} \, \langle p | \, \bar{\psi}(0,w^-,{\bf 0}_t) \gamma^+ V_{[0,w]} (u_{\rm J}) \,  \psi(0) | p \rangle. 
\end{equation}
Our symbol ``${\rm F.T.}$'' is a short-hand for the 
Fourier transform from coordinate space to momentum space.
The basic structure of the definition is evidently that of a number density; a 
quark is extracted from the proton state at position $0$ and propagates 
to position $(0,w^-,{\bf 0})$.  A few subtleties should be noted, however.  
One is that the definition contains UV divergences which must be renormalized.
This gives dependence on an extra scale $\mu$, and ultimately results in the 
well-known DGLAP evolution equations for the integrated PDF.  The other is that, for 
a gauge invariant definition, the PDF must contain a path ordered exponential 
of the gauge field that connects the points $0$ and $(0,w^-,{\bf 0}_t)$.
This is the gauge link and its formal definition is
\begin{equation}
\label{eq:wilsonline}
V_{[0,w]} (u_{\rm J}) = P \exp \left(-igt^a \int_{0}^{w^-} \, d \lambda \,  u_J \cdot A^a(\lambda u_J) \right).
\end{equation}
The path of the gauge link is determined by the 
light-like vector $u_{\rm J} = (0,1,{\bf 0}_t)$.
That is, the gauge link follows a straight path connecting $0$ 
and $(0,w^-,{\bf 0}_t)$ along the exactly light-like minus direction.
In Feynman graph calculations, the contribution 
from the gauge link corresponds to the so-called ``eikonal factors,'' 
which have definite Feynman rules that 
follow naturally from factorization proofs.  
After a sum over graphs, and the application of appropriate approximations and Ward identity arguments, 
extra collinear gluons like those shown in figure~\ref{fig:basicfact}(a) for SIDIS factor into gauge link contributions.
In figure~\ref{fig:basicfact}(b),
the eikonal factors are shown as gluon attachments from the target-collinear bubble to a double line.
\begin{figure*}
\centering
  \begin{tabular}{c@{\hspace*{5mm}}c}
    \includegraphics[scale=0.6]{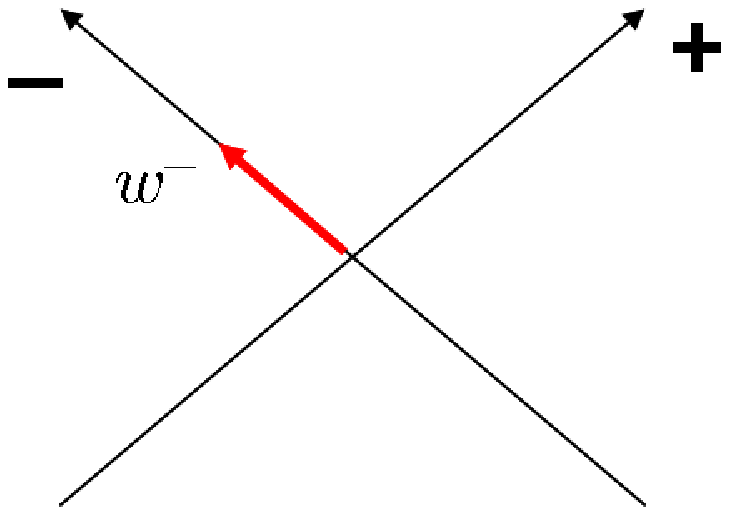}
    &
    \includegraphics[scale=0.6]{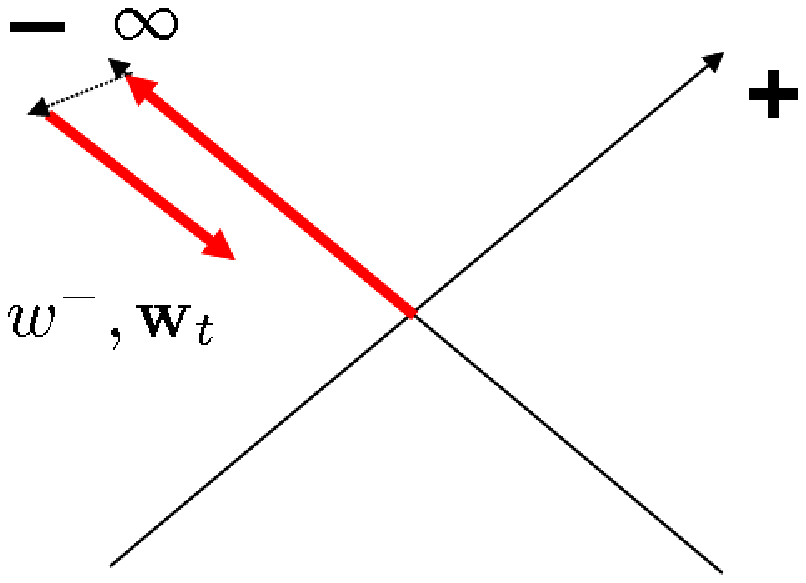}
  \\
  (a) & (b)
  \end{tabular}
\caption{(a) Simple light-like gauge link structure in integrated the PDF.  (b) First try at a gauge link structure for the TMD PDF.
In both of these diagrams, the thick red arrows represent the main light-like legs of the gauge link.  In (b), the then dotted link at connecting 
the main legs at light-cone 
minus infinity points in the transverse direction, which is perpendicular to the page.}
\label{fig:gaugelinks1}
\end{figure*}

To aid in the discussions of how the gauge link should be 
modified in more complicated circumstances, it is useful to 
visualize the coordinate space geometry of the gauge link 
structure.  For example, for the integrated PDF in equation~(\ref{eq:intDEF}), 
the gauge link follows a straight path in the exactly light-like minus 
direction, as illustrated by the Minkowski-like diagram in figure~\ref{fig:gaugelinks1}(a).

\section{TMD definitions}
\label{sec:TMDdefs}

The most natural first try at extending the PDF definition 
in equation~(\ref{eq:intDEF}) to the TMD case is to simply 
leave the integration over transverse momentum in the TMD PDF 
definition undone.  That is, instead of Eq.~(\ref{eq:intDEF}) one may try
\begin{equation}
\label{TMDdef1}
\Phi(x,{\bf k}_t) =  {\rm F.T.} \, \langle p | \, \bar{\psi}(0,w^-,{\bf w}_t) \gamma^+ U_{[0,w]}(u_J) \psi(0) \, | p \rangle.
\end{equation}
The separation is now $0$ and $(0,w^-,{\bf w}_t)$ --- it has acquired 
a transverse component and the Fourier transform is now 
in both $w^-$ and ${\bf w}_t$.  As a result, the structure of the 
gauge link $U_{[0,w]}(u_J)$ must also be modified from the 
simple straight light-like $V_{[0,w]} (u_{\rm J})$ gauge link of equation~\ref{eq:intDEF}.  
The eikonal attachments on either side of the cut in figure \ref{fig:basicfact}
still give minus-direction Wilson 
lines, but now in order to have a closed link there must 
also be a small transverse detour at light-cone infinity.
This detour arises naturally from boundary terms that are needed as subtractions
to make higher twist contributions gauge invariant~\cite{gaugelinks2}.
The first try at a coordinate space picture of the gauge link should therefore 
be more like the hook shaped line shown in figure~\ref{fig:gaugelinks1}(b).

\begin{figure*}
\centering
\includegraphics[scale=.6]{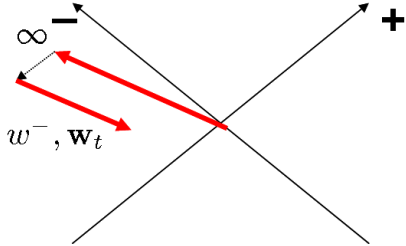}
\caption{Gauge link structure for the TMD PDF tilted away from the exactly light-like direction.}
\label{fig:tiltedlink}
\end{figure*}

The gauge link structure in equation~\ref{TMDdef1}, with its two exactly light-light 
legs and a transverse link at infinity is commonly 
cited as the gauge link that is necessary for the 
definition of the TMD PDFs.  However, there are a number of further subtleties, and we will find that 
the definition needs to be modified.
One complication is that
rapidity divergences, which in collinear factorization would cancel in the sum of graphs, remain uncanceled in the 
definition of the TMD correlation functions.  
Rapidity divergences correspond to gluons moving with 
infinite rapidity in the direction opposite the containing hadron, and remain even when infrared gluon mass 
regulators are included.  
(For a more complete review of these and related issues, see for example~\cite{collrevs}.)  
The most common way to regularize the light cone
divergences is to make the gauge links slightly non-light-like.  
In the coordinate space picture, the gauge link therefore becomes more like the 
\emph{tilted} hook shape in figure~\ref{fig:tiltedlink}.  This introduces a new arbitrary rapidity 
parameter -- the ``tilt'' of the gauge link.  A generalization of 
renormalization group techniques is needed to recover predictability in the factorization 
formula.  A system of evolution equations for the TMD 
case was developed by Collins, Soper and Sterman (CSS) and has 
been successfully applied to specific processes~\cite{CSS}.

A complete treatment of TMD-factorization involves soft gluons, which 
give rise to an extra ``soft factor'' $S({\bf q})$ 
in the factorization formula of equation~(\ref{eq:basic}).  The TMD-factorization formula then becomes
\begin{equation}
\label{eq:basic2}
d \sigma \sim | \mathcal{H} |^2 \otimes \Phi(x,{\bf k}_T) \otimes \Delta(z,{\bf k}^\prime_T) \otimes S({\bf h}_T) \;  
\delta^{(2)} ({\bf q}_T + {\bf k}_T - {\bf k}^\prime_T - {\bf h}_T).
\end{equation}
The soft factor describes the role of gluons with 
nearly zero center-of-mass rapidity.  One difficulty
with the usual presentation of the CSS formulation is 
that the explicit appearance of a soft factor seems 
somewhat counter to the basic parton model intuition 
wherein all non-perturbative effects are associated
with functions for each external hadron with simple and specific probabilistic interpretations.  
A natural hope is that, with an appropriate sequence of redefinitions, 
the role of the soft gluons can be absorbed into 
the definitions of the PDFs and FFs.  The recent work of 
Collins~\cite{collins} has shown how this is possible.  Indeed, this treatment of the soft factor 
is necessary for a completely correct factorization derivation 
with fully consistent definitions for the correlation functions.

While the CSS formalism has been implemented for 
specific spin independent processes (see, for example,\cite{Landry:2002ix}) , much work remains to be done 
in tabulating and classifying the TMDs.  This is especially true for cases that involve spin.
Work in this direction has been started in~\cite{tedmert}.

\section{TMD-factorization breaking}
\label{sec:nofact}

\begin{figure*}
\centering
  \begin{tabular}{c@{\hspace*{5mm}}c}
    \includegraphics[scale=0.6]{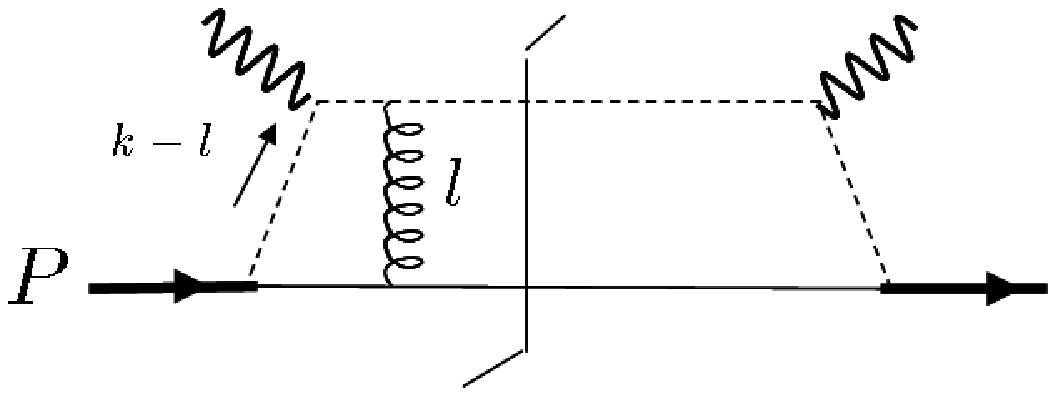}
    &
    \includegraphics[scale=0.6]{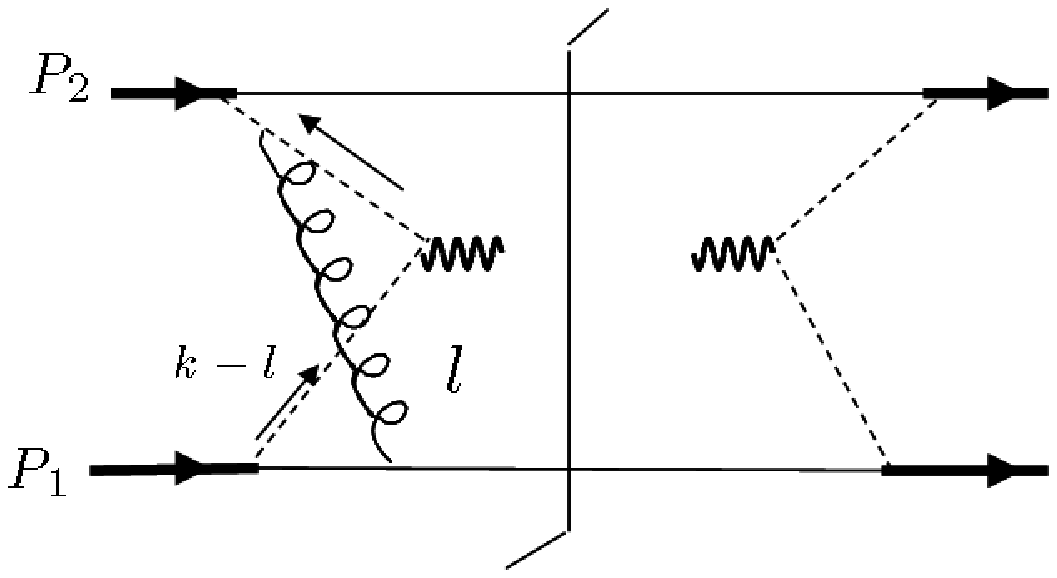}
  \\
  (a) & (b)
  \end{tabular}
\caption{A single collinear gluon attachment (a) after the hard scattering in SIDIS and (b) before the hard scattering in DY.}
\label{fig:disdy}
\end{figure*}

The discussion of the last few sections has focused on 
situations where factorization is known to hold.
There are also, however, situations where TMD-factorization is 
now known to break down~\cite{gaugelinks,collins.qiu,Collins:2007jp,Bomhof:2007xt,Rogers:2010dm}.  The key issue is the failure of the usual 
Ward identity arguments that ordinarily allow eikonalized 
gluons to be factorized and identified with a particular
gauge link structure in the definitions of the TMDs.  A hint of 
what leads to TMD-factorization breaking is already suggested by the 
well-known overall relative sign flip in the Sivers function for SIDIS as 
compared to the Drell-Yan (DY) process~\cite{sineflip}.  The difference comes 
because in the SIDIS TMD-factorization formula, the gauge link in 
the Sivers function (a spin-dependent TMD PDF) is future pointing, 
whereas it is past pointing in the the DY case.  At the level of 
Feynman graphs, the difference can be seen in the fact that the 
``extra'' gluons which contribute to the gauge link attach before 
the hard scattering in one case, and after the hard scattering in the other
(see figure~\ref{fig:disdy}).  This illustrates that the direction of the flow of color through 
the eikonal lines is 
a critical factor in the definition of the correlation functions.

\begin{figure*}
\centering
  \begin{tabular}{c@{\hspace*{5mm}}c}
    \includegraphics[scale=0.6]{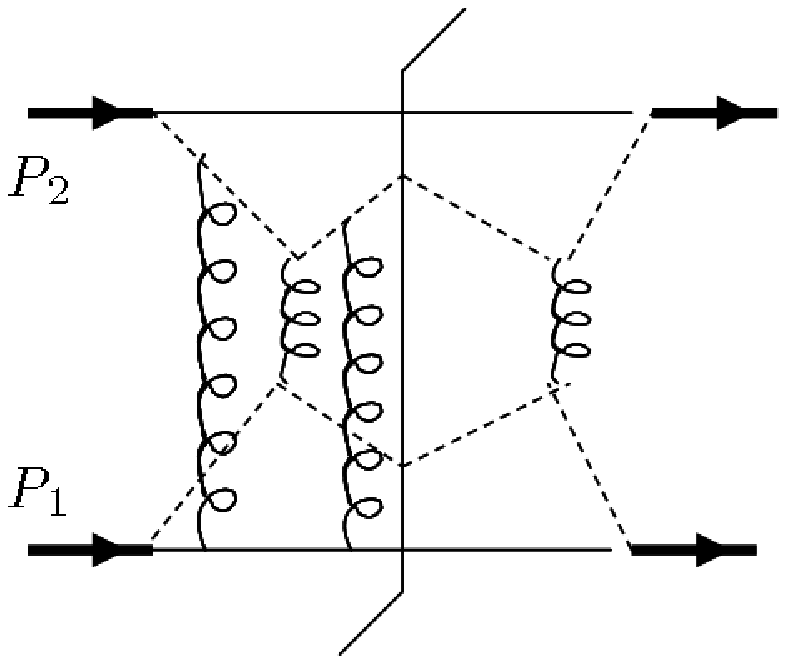}
    &
    \includegraphics[scale=0.6]{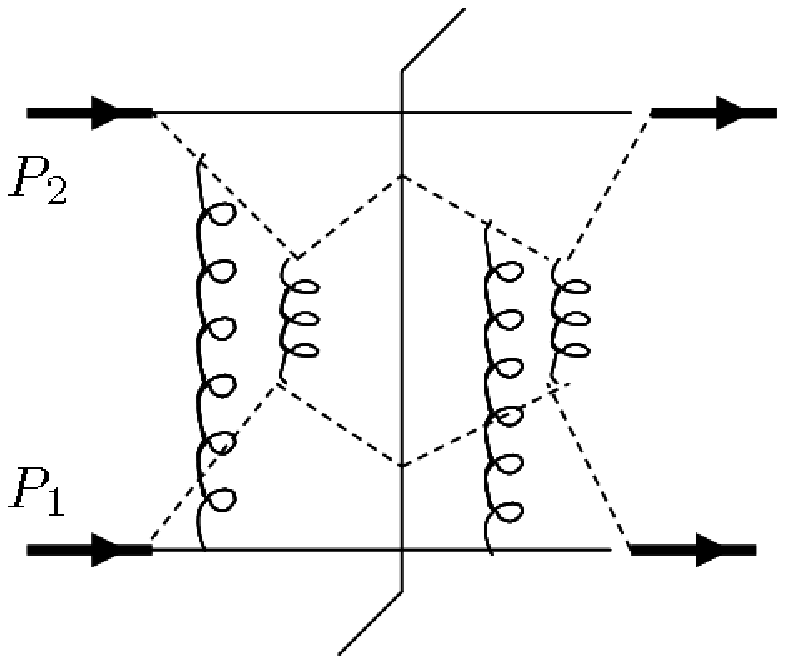}
  \end{tabular}
\caption{Graphs with extra collinear gluons attaching at the ``wrong'' side of the graph 
for them to be identified with the 
standard universal gauge link 
contributions in the lower proton PDF.}
\label{fig:nonfactor}
\end{figure*}

In the more complicated hadro-production processes, $H_1 + H_2 \to H_3 + H_4 + {\rm X}$, where $H_3$ and $H_4$ may 
be either jets or hadrons, a reasonable first approach would be to traced 
the flow of color through the eikonal factors and use analogous arguments 
to what we used for SIDIS and DY in the previous section. 
Looking at graphs like figure~\ref{fig:nonfactor}, one finds that the resulting structures  
are not simply the future or past pointing gauge links familiar from SIDIS or DY, but 
rather are complicated and highly process dependent objects~\cite{gaugelinks}.
That this corresponds (at least) to a breakdown 
of universality is most directly seen in an explicit 
spectator model calculation.  For example, one may consider an Abelian scalar-quark / Dirac spectator model 
with multiple flavors as in~\cite{collins.qiu}.  
Then, in addition to the standard gauge link attachments, there are extra gluon attachments that do not cancel
in a simple Ward identity argument, and which give contributions that are not consistent with having a simple gauge link like what is
found SIDIS (figure~\ref{fig:tiltedlink}) or DY (opposite pointing).  

\begin{figure*}
\centering
\includegraphics[scale=.6]{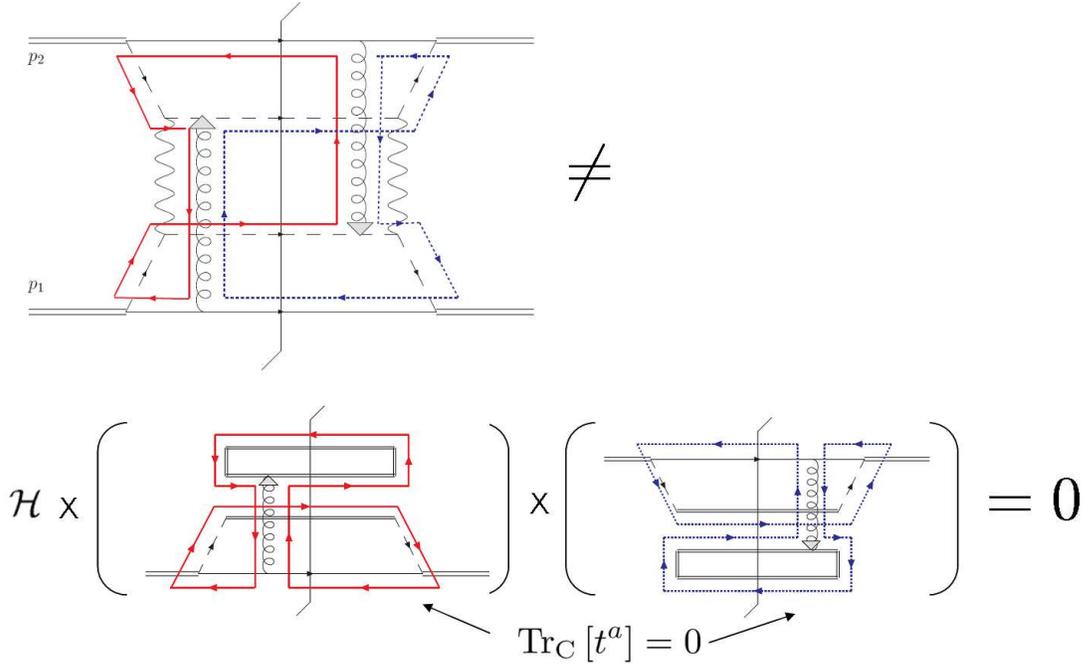}
\caption{A sample graph that illustrates the entanglement of color that leads to a failure of factorization.
The overall color factor of the first graph is non-zero, but the product of two one-gluon contributions 
to the matrix elements with Wilson loops is exactly zero.
}
\label{fig:factviolation}
\end{figure*}

Therefore, it is clear that there is \emph{at least} a violation of universality in 
the hadro-production of hadrons.  The natural next approach to try is to maintain a 
basic factorization structure, but to loosen the requirement that the TMDs be universal, resulting in  a 
kind of ``generalized'' TMD-factorization formalism.  That is, the cross section might still be expected 
to factorize order-by-order into a hard part and well-defined, albeit non-universal, matrix elements for 
each separate external hadron~\cite{Bomhof:2007xt}.
However, a careful order-by-order consideration of multiple gluons in the 
derivation of TMD-factorization shows that even this is not possible~\cite{Rogers:2010dm}.  
If, for example, one extends the model of~\cite{collins.qiu} to allow 
the gluons to carry color (while still considering a hard part that involves only the exchange
of a colorless boson) then it is straightforward to see that the flow of color spoils the possibility 
of factorizing the graph into TMD PDFs with separate gauge links for each TMD, regardless
of what kind of gauge link geometries are allowed.
(See figure~\ref{fig:factviolation} for an illustration.)
Therefore, the problem with factorization in the hadro-production of hadrons 
is more than just a problem with universality -- separate correlation 
functions cannot even be defined in a way that is consistent with factorization.
 
Because the graphs in~\ref{fig:factviolation} illustrate the failure of factorization in terms of a failure of color flow to factorize,
it is tempting to conclude that the TMD-factorization breaking is purely an issue of complicated color flow.  
This is an oversimplification of the issue, however.  The root of the problem is a failure of Ward identity arguments, 
which normally allow ``extra'' gluons to be factorized after a sum over graphs.  The Ward identity arguments are only 
valid after an appropriate sequence of contour deformations on the momentum integrals.  In the case 
of hadro-production of hadrons the necessary deformations 
are prohibited.  In other cases where the direction of color flow may at first 
appear to pose a problem for factorization (such as in $e + p \to h_1 + X$ and $e + p \to h_1 + h_2 + X$), the necessary contour 
deformations are possible and factorization holds.  (See the explanation in chapter 12 of~\cite{collins}.)   

\section{Conclusions}

We have discussed the basic status of issues related to the definitions of TMDs with 
a focus on the complications that can arise when determining the gauge link structures 
that are consistent with factorization.  We have also described the 
breakdown of TMD-factorization in the case of hadro-production of hadrons.
To summarize, we list the status of TMD-factorization for various well-known processes with a check mark 
for processes where factorization appears to be valid and $!!$ where it has been shown to fail:

\begin{center}
\renewcommand{\labelitemi}{$\checkmark$}
\begin{itemize}
\item Semi-inclusive deep inelastic scattering ($e^+ + p \to h_1 + X$).
\end{itemize}
\renewcommand{\labelitemi}{$\checkmark$}
\begin{itemize}
\item Drell-Yan (up to overall minus signs for some spin-dependent TMDs).
\end{itemize}
\renewcommand{\labelitemi}{$\checkmark$}
\begin{itemize}
\item Back-to-Back hadron or jet production in $e^+ e^-$ annihilation.
\end{itemize}
\renewcommand{\labelitemi}{$\checkmark$}
\begin{itemize}
\item Back-to-back hadron or jet production in DIS ($e^+ + p \to h_1 + h_2 + X$).
\end{itemize}
\renewcommand{\labelitemi}{$!!$}
\begin{itemize}
\item Hadro-production of back-to-back jets or hadrons ($H_1 + H_2 \to H_3 + H_4 + X$).
\end{itemize}
\end{center}

In cases where TMD-factorization is valid, there is still much work left to be done 
(and much potential insight to be gained) in terms of implementing the evolution of precisely 
defined TMDs~\cite{tedmert}.  Much already exists for the case of unpolarized scattering, but even here
the most complete and formal identification of evolution effects with separate TMDs has only recently been 
clarified in~\cite{collins}.  For polarization dependent functions, it is also important to include evolution, but
to date there has been very little work that accounts for evolution in actual fits to data.

Finally, the experimental search for TMD-factorization breaking effects opens the possibility of 
new and exciting insights into the transverse dynamics of hadronic collisions.  The breakdown of TMD-factorization in 
the hadro-production of hadrons implies that unexpected and exotic correlations between partons in \emph{different} hadrons 
can exist.
Calculations that allow for experiments to distinguish between factorization and factorization-breaking scenarios
are therefore very important, and a quantitative understanding of factorization (via the methods of~\cite{Dominguez:2010xd}, for example) are 
part of the next step toward understanding hadronic structure in high energy collisions. 

\subsection*{Acknowledgments}
We would like to thank the organizers of this stimulating workshop for their invitation and hospitality and the 
Institute for Nuclear Theory for financial support.
This research is part of the research program of the ``Stichting voor Fundamenteel Onderzoek der Materie (FOM)'', 
which is financially supported by the ``Nederlandse Organisatie voor Wetenschappelijk Onderzoek (NWO)'';
it is also part of the FP7 EU-programme Hadron Physics (No. 227431).

\printindex

\end{document}